\documentclass[aps,prl,twocolumn,groupedaddress,showpacs]{revtex4}
\usepackage{graphicx}

\begin{document}

\title{Photonic band structure of ZnO photonic crystal slab laser}
\author{A. Yamilov, X. Wu, and H. Cao}
\email{a-yamilov@northwestern.edu}
\affiliation{Department of Physics and Astronomy, Northwestern University, Evanston, IL, 60208}

\begin{abstract}
We recently reported on the first realization of ultraviolet photonic crystal laser based on zinc oxide [Appl. Phys. Lett. {\bf 85}, 3657 (2004)]. Here we present the details of structural design and its optimization. We develop a computational super-cell technique, that allows a straightforward calculation of the photonic band structure of ZnO photonic crystal slab on sapphire substrate. We find that despite of small index contrast between the substrate and the photonic layer, the low order eigenmodes have predominantly transverse-electric (TE) or transverse-magnetic (TM) polarization. Because emission from ZnO thin film shows strong TE preference, we are able to limit our consideration to TE bands, spectrum of which can possess a complete photonic band gap with an appropriate choice of structure parameters. We demonstrate that the geometry of the system may be optimized so that a sizable band gap is achieved. 
\end{abstract}

\maketitle


\section{I. Introduction}

A photonic crystal slab (PhCS) is a layer of dielectric where refractive index is periodically modulated within the plane of the slab\cite{phcs_old1,phcs_old2,Johnson:1999}. Such devices have attracted much attention because of their potential applications to various optoelectronic devices and circuits \cite{Krauss:1996,Painter:1999,noda_phcs_filter,Chow:2000,Akahane:2003,phcs_applications,soukoulis}. PhCS may lead to a complete photonic band gap for the guided modes when the wave propagation in all in-plane directions is forbidden due to Bragg interference. Moreover, its planar geometry makes it easy to incorporate on a chip, as well as offers a possibility of low-threshold\cite{Painter:1999,Imada:1999,Benisty:1999,Hwang:2000,low_threshold_phcs,bandedge_exp} electrically-driven compact laser source\cite{park_electrically_pumped}. Defect cavities in these photonic devices can have high quality factor and small modal volume\cite{Johnson:2001b,Vuckovic:2001,Srinivasan:2002,Akahane:2003}. So far the experimental efforts mainly concentrated on PhCS made of III-V semiconductors.\cite{Painter:1999,Imada:1999,Benisty:1999,Hwang:2000,low_threshold_phcs,bandedge_exp,park_electrically_pumped} They operate in the infrared (IR) communication frequencies (see also \cite{notomy_visible_phcs}), where the structural feature size is on the scale of $0.25\mu m$.

There is technological and commercial demand for  compact and integrable laser sources in near the ultraviolet (UV) range of optical spectrum. Recently, we reported the first realization of UV PhCS laser that operates at room temperature\cite{our_apl}. This required overcoming several experimental and design challenges. 

First of all, because of the shorter wavelength, PhCS requires smaller structural features with sub $100nm$ size. The fabrication of such miniature structures is technologically challenging for commonly used wide band gap semiconductors such as GaN and ZnO. Compared with the other wide band gap materials ZnO, has the advantage of large exciton binding energy ($\sim 60$ meV), that allows efficient excitonic emission even at room temperature\cite{Park:1966}. Using Focused Ion Beam (FIB) etching technique\cite{our_apl} we were able to achieve more than a four-fold reduction, compared to IR PhCS, of the feature size in ZnO (see also \cite{xiaohua_exp_study}). 

Secondly, high quality ZnO films are usually grown on the lattice matched sapphire substrate\cite{xiang}, that results in relatively low refractive index contrast between PhCS and substrate. Furthermore, as we detail below, the substrate-PhCS-air rather then air-PhCS-air geometry in vertical direction disallows the classification of the photonic modes into symmetric and antisymmetric with respect to reflection in the mid-PhCS plane\cite{Johnson:1999,Chow:2000,qiu_asymetric}. Separation of the guided modes according to their symmetry in IR PhCS is usually ensured due to ``undercutting'', i.e. removing the material directly below the photonic layer\cite{Painter:1999,bandedge_exp,park_electrically_pumped}. The separation can also be approximately\cite{Chow:2000,phcs_on_low_index_substrate} done for systems with a large index contrast between the substrate and PhCS. Neither option is readily available for ZnO-based structures, due to its material properties. At first glance, this makes photonic band gap (PBG) unattainable in our system. The purpose of this paper is to design ZnO-based PhCS, which exhibits PBG effects.

\begin{figure}
\centerline{\rotatebox{0}{\scalebox{0.5}{\includegraphics{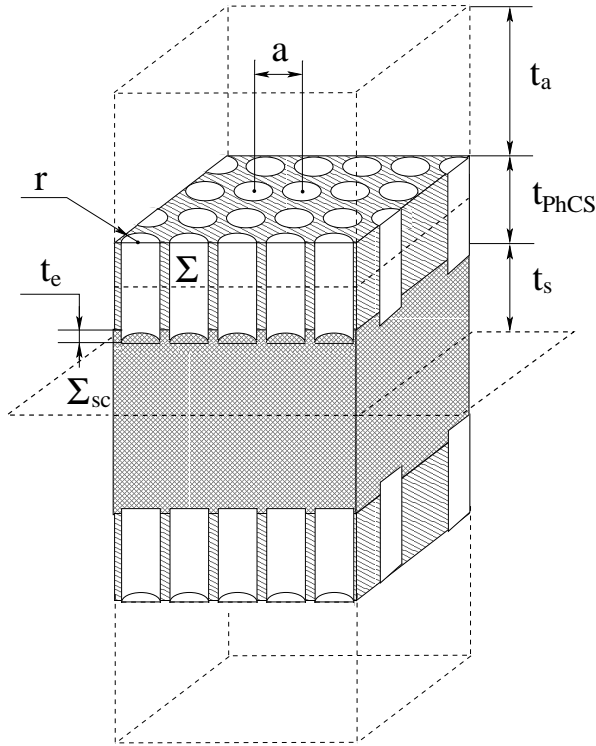}}}}
\caption{\label{geometry}  Schematic diagram of the vertical super-cell used in photonic band structure simulation. Cylindrical air holes of radius $r$ are arranged in a hexagonal pattern with lattice constant $a$. The holes extend throughout the slab of thickness $t_{PhCS}$ and may penetrate into the substrate to the depth of $t_e$. The thicknesses $t_a$ and $t_s$ need to be chosen sufficiently large (see text for discussion). The cell is symmetric with respect to reflection in $\Sigma_{sc}$ plane. For convenience we also introduce a middle plane of the slab, $\Sigma$. }
\end{figure}

In Section II we introduce a computational super-cell technique that allows the calculation of photonic band structure of PhCS on a dielectric substrate. Then we present the detailed analysis of symmetry and polarization of the photonic modes belonging to different bands. We show that low-order photonic bands can still be approximately classified as TE and TM. In Section III we make use of the experimental fact that PhCS is made of single-crystalline hexagonal ZnO thin film, which is grown along c- crystallographic axis\cite{xiang}. Since the polarization of exciton emission in the ZnO thin film is predominantly perpendicular to the c-axis\cite{te_gain,random_laser}, the emission goes preferably into TE-polarized photonic modes. Coincidentally, the air-holes-in-a-dielectric PhCS configuration can possess complete PBG (with certain parameters) for namely TE polarized modes. Therefore, we consider only TE-polarized photonic modes in our band structure calculation. Finally, in Section IV we systematically study the position and the size of PBG as a function of geometrical parameters: air-hole radius $r$, slab thickness $t_{PhCS}$ and etch depth $t_e$ as illustrated in Fig. \ref{geometry}. We conclude our paper with a discussion section.

\section{II. Super-cell photonic band structure calculation}

Although ``Bragg reflection'' plus ``index guiding'' description of the electromagnetic field in PhCS can elucidate many physical phenomena that occur in such system, it is naive because both in-plane and vertical propagation or confinement are inseparable. In order to obtain the photonic modes of PhCS one needs to solve the problem self-consistently within the same framework. This can be done with a plane-wave expansion method\cite{Johnson:2001} that {\it de facto} became a standard in the field. 

PhCS designed for IR applications have an inherently large size of the structural units owing to relatively long wavelengths involved. In these systems it is possible to fabricate PhCS cladded with air above and below (see e.g. \cite{Painter:1999,park_electrically_pumped,bandedge_exp}). This design is advantageous for a number of reasons. First, this is  a symmetric structure -- there exists a reflection symmetry with respect to the plane that goes through the middle of PhCS. This symmetry allows two orthogonal classes of eigenfunctions of Maxwell equations: symmetric and antisymmetric\cite{Johnson:1999}. Super-cell for the band-structure calculation can be easily constructed by repeating ...-air-PhCS-air-PhCS-... indefinitely in the vertical direction\cite{Johnson:1999}. It was shown that low-lying (E-field) symmetric modes are predominantly TE polarized, while antisymmetric ones are TM polarized. Due to superior guiding properties of TE modes and the fact that self-supporting membrane percolated by air holes (unlike collection of dielectric cylinders in air) exhibits PBG for the same polarization, this geometry is widely used in practice. The super-cell photonic band structure calculations also give unphysical modes that do not correspond to the guided modes of PhCS (confined to the slab). Fortunately, physical (guided) modes can be separated using the concept of light-cone as follows. In the result of photonic band structure calculation one obtains the frequency $\omega$ for a given in-plane wavevector ${\bf k_{||}}$. This EM mode (concentrated inside the photonic layer) is coupled only to the modes outside the layer with the same ($\omega$,${\bf k_{||}}$). If $\omega<ck_{||}$, then $k_z=\sqrt{\omega^2/c^2-k_{||}^2}$ is imaginary outside PhCS. Thus the light intensity decays exponentially in the z-direction away from the layer. This describes a guided mode. In the opposite case, $k_z$ is real and light can escape to the infinity. Such a mode is a leaky mode, which cannot be accurately described in the super-cell calculations. A continuum of modes with $\omega>ck_{||}$ form a light cone (a cone in $\omega-k$ space). Therefore, PBG obtained in all planar geometries are, strictly speaking, photonic band gaps for guided modes.

In our UV PhCS laser \cite{our_apl} we used ZnO because of its wide electronic band gap and robustness of the optical properties with respect to damage caused by FIB. There are two consequences relevant for the band structure calculation. First, in order to produce high quality films, ZnO was grown epitaxially on sapphire {\it substrate with large refractive index} $n_s=1.78$. Second of all, our choice of the materials and miniature structure dimensions, do not currently allow for removal of the underlying layer of sapphire --- {\it ``undercutting'' is impossible}. These experimental limitations lead to severe restrictions. (i) Presence of the sapphire substrate lifts the reflection symmetry\cite{Johnson:1999}. This results in mixing between TE-like and TM-like classes of electromagnetic modes. One may expect that this effect should be large in view of relatively small index contrast between sapphire and ZnO ($n_{ZnO}=2.35$). Below we will show that this expectation is not completely borne out. (ii) The most stringent restriction is imposed by the presence of the second light cone -- that reflects the possibility of radiative escape into the sapphire substrate. Indeed, although light leakage into the air occurs for $\omega>ck_{||}$, leakage into the substrate starts already for $\omega>(c/n_s)k_{||}$. Avoiding this problem requires the use of high filling fraction in ZnO PhCS ($f>50\%$), that in turn drives down the size of the features to be produced in the experiment. 

\begin{figure}
\vskip 0.25in
\centerline{\rotatebox{0}{\scalebox{0.35}{\includegraphics{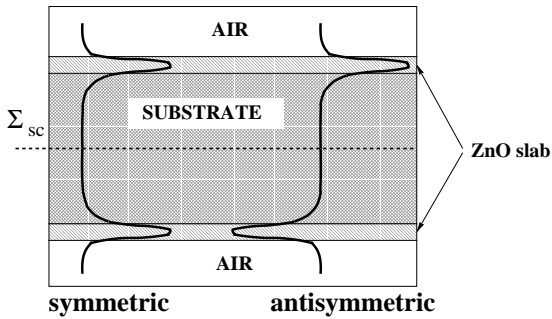}}}}
\caption{\label{mode_symmetry} Reflection symmetry allows for solutions symmetric and antisymmetric with regard to reflection in $\Sigma_{sc}$. This artificial symmetry helps to separate the guided modes from the leaky ones, see text for details.}
\end{figure}

A band structure calculation technique for the asymmetric geometry like ours was proposed in Ref. \cite{Johnson:1999}. The authors of Ref. \cite{Johnson:1999} proposed to use a super-cell where ...-substrate-PhCS-air-substrate-PhCS-air-.. is periodically repeated in z-direction. Special care needs to be exercised in order to avoid unphysical solutions that correspond to the guided modes in the substrate layers bounded by air and a PhCS. In our simulation we constructed a different super-cell by combining the structure and its mirror image (shown in Fig. \ref{geometry}) for the photonic band  calculation\cite{Johnson:2001}. Our choice of the unit cell is motivated by the following. Due to particular arrangement of the layers in our super-cell, there exists an artificial symmetry -- reflection plane $\Sigma_{sc}$ in Fig. \ref{geometry}. Therefore, there is an artificial classification of the modes into symmetric and antisymmetric, as it is schematically shown in Fig. \ref{mode_symmetry}. This property provides a way to control the precision of the band structure calculation. Indeed, only for the PhCS guided modes (with eigenfrequencies outside of air and substrate light cones), and only when thicknesses of air ($t_a$) and substrate ($t_s$) layers are chosen to be sufficiently large, the eigen-frequency of $\Sigma_{sc}$-symmetric and $\Sigma_{sc}$-antisymmetric modes are nearly degenerate. This can be seen from Fig. \ref{mode_symmetry}: there should be no difference in eigenenergy between the functions when the value of the eigenfunction is close to zero at (i) $\Sigma_{sc}$ plane and (ii) the top/bottom of the super-cell. The second condition arises because of the periodicity in the vertical direction --- the top/bottom of the super-cell is also a middle plane that separates two closest PhCS layers. 

Furthermore, one may think that the doubling of the cell size should increase the computation time. However, at least for a PhCS with a hexagonal symmetry, $\Sigma_{sc}$ contains a center of inversion. The latter allows to perform the photonic band structure calculation assuming inversion symmetry that shortens the calculation time (of one symmetry) by a half. In all simulations reported below, we checked the consistency of the results with the procedure outlined above, then studied the band structure of the modes outside of the substrate (more restrictive) light cone. 
\begin{figure}
\hskip -1in
\centerline{\rotatebox{0}{\scalebox{0.5}{\includegraphics{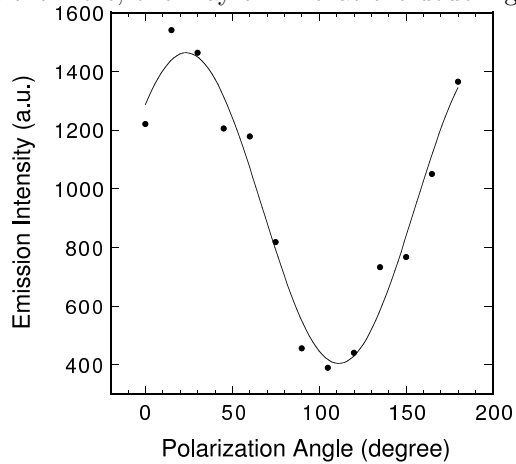}}}}
\vskip -1.7in
\caption{\label{polarization_exp} Experimentally measured emission intensity from $600 nm$ ZnO film as a function of polarization angle. The maximum intensity at about 25$^o$ corresponds to E-field polarization parallel to the plane of the film (TE polarization).}
\end{figure}
\begin{figure}
\centerline{\rotatebox{-90}{\scalebox{0.33}{\includegraphics{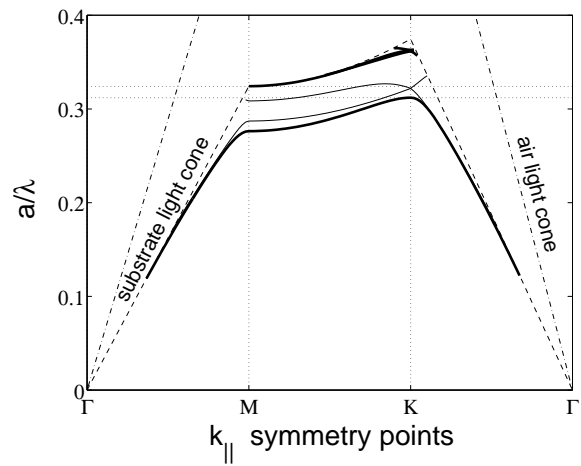}}}}
\caption{\label{cal} Calculated band structure of ZnO photonic crystal slab air cylinder radius $r/a=0.25$, and slab thickness $t/a=1.4$. The refractive indexes for ZnO and sapphire are 2.35 and 1.78. Thin and thick lines represent TM and TE polarized modes respectively.}
\end{figure}

\begin{widetext}
\begin{figure}
\hskip 5in
\rotatebox{-90}{\scalebox{0.65}{\includegraphics{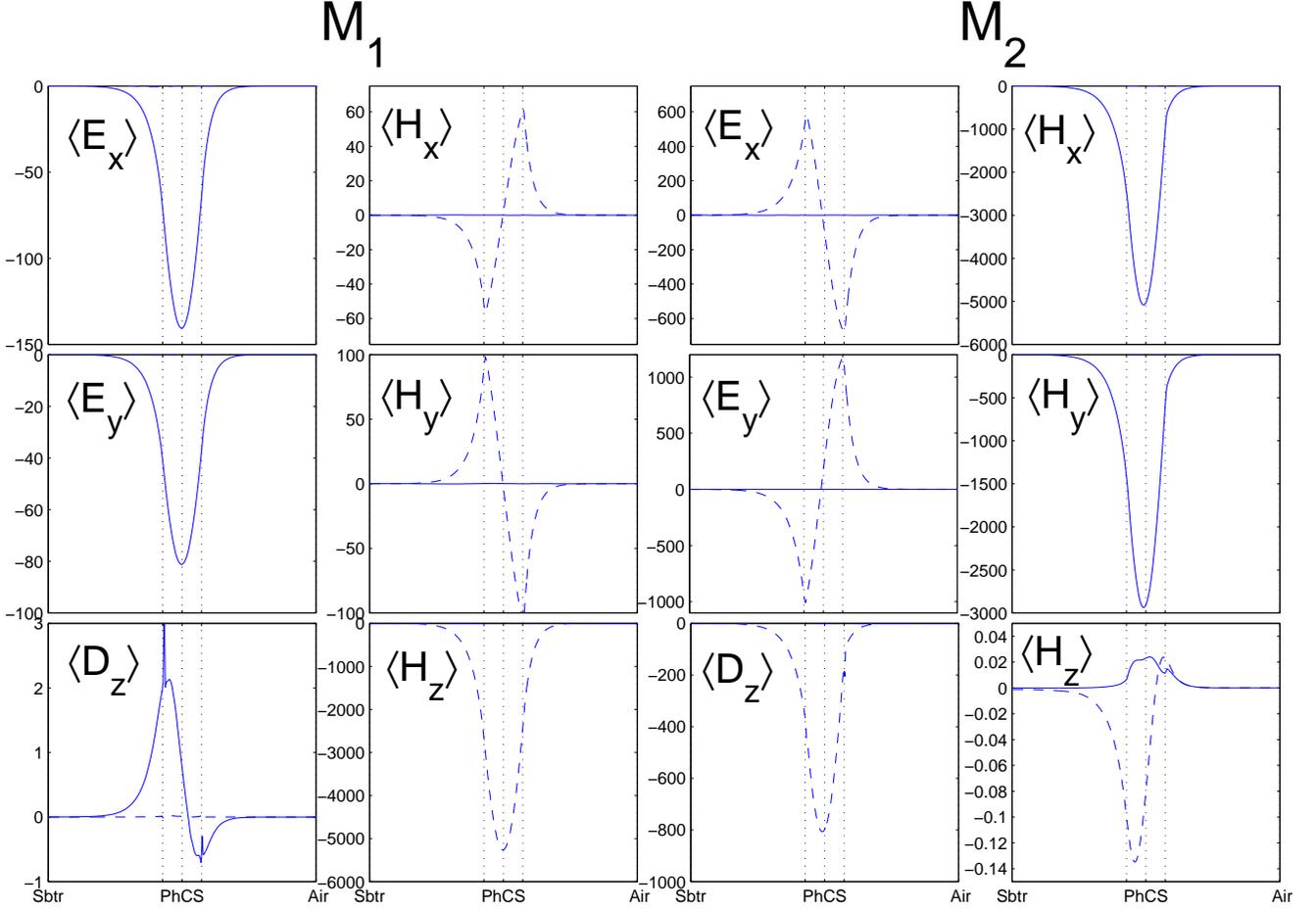}}}
\caption{\label{fields} Integrated according to Eq.(\ref{averaged_field}) $x,y$ and $z$ components of the electric, $E$, and magnetic, $H$, field. Solid/dashed lines represent real/imaginary part of the field. First/second two columns correspond to the first/second mode of Fig. \ref{cal} taken for $k_{||}$ at $K$ symmetry point. The same normalization is used for all field components, so that their values can be compared. Mode $M_1/M_2$ shows a pronounced TE/TM polarization.}
\end{figure}
\end{widetext}

\section{III. Photonic mode classification: Polarization}

We measured the polarization of ZnO emission intensity from a $600 nm$-thick ZnO film on sapphire substrate. c-axis of ZnO is perpendicular to the surface of the film. ZnO is optically excited by the third harmonic  of a mode-locked Nd:YAG laser ($355nm$, $10Hz$, $20ps$). A cylindrical lens ($f=100 mm$) is used to focus the pump beam to a strip normal to the edge of the sample. The emission was collected at $1\mu J$ pump pulse energy and the pump area was about $2mm\times 50\mu m$.  Fig. \ref{polarization_exp} clearly demonstrates that the emission is strongly polarized with the electric field parallel to the film (TE polarization) \cite{random_laser,te_gain}. Because the emission in the ZnO thin film is predominantly TE polarized we need to consider consider only the TE-polarized photonic modes. This motivates a detailed analysis of the polarization of the modes in the experimentally relevant PhCS.

In Fig. \ref{cal} we show an example of calculated photonic band structure. Air light-cone boundary, $\omega=ck_{||}$, and substrate light cone boundary, $\omega=(c/n_s)k_{||}$, are shown with dash-dot and dashed lines respectively. The presence of sapphire cone limits any bandgap to $a/\lambda<0.325$. For PhCS laser operation we need to overlap a photonic band gap with gain spectrum of ZnO, $380nm<\lambda<400nm$. Consequently, the lattice constant in the hexagonal air-hole pattern of PhCS has an upper bound of $a<130nm$. To fabricate such small features we used the FIB etching\cite{our_apl}.

Since the presence of the substrate removes the symmetry with respect to reflection in PhCS middle plane ($\Sigma$ in Fig. \ref{geometry}), one cannot separate the modes into two independent classes and consider them separately. This separability was crucial for obtaining sizable PBG in a photonic membrane (or strictly two-dimensional structures). For comparison, we calculated the band structure of the system with the same parameters as in Fig. \ref{cal} but sapphire substrate was replaced by air. The obtained dispersions looked qualitatively similar to the ones in Fig. \ref{cal}, so that it was possible to make band-to-band correspondence. This has been used previously to justify the existence of PBG in the spectrum of TE modes\cite{Chow:2000}. In our case the refractive index of the substrate is quite large, moreover it may even become comparable to the effective index of PhCS as the patterning of ZnO film with original refractive index $n_{ZnO}=2.35$ will reduce it to just a little above that of the sapphire. These considerations warrant an in-depth analysis of the mode polarizations.

In order to determine the degree of polarization of the electromagnetic eigenmodes of PhCS, we studied 
\begin{equation}
\langle F(z)\rangle=\int F(x,y,z)\;dx\;dy,
\label{averaged_field}
\end{equation}
where $F$ represents $x,y$ or $z$ component of electric or magnetic field. Fig. \ref{fields} shows these quantities for the first two bands taken at $M$-symmetry point of Fig. \ref{cal}. It is clearly seen that the first (the lowest frequency) mode $M_1$ is strongly TE polarized: $E_x$, $E_y$ and $H_z$ are more than one order of magnitude larger the other components. Furthermore, as discussed above dominant components are nearly symmetric with respect to middle plane of PhCS, whereas $H_x$, $H_y$ and $E_z$ are close to antisymmetric. The deviation from the perfect symmetry is entirely due to the presence of the sapphire substrate. Comparing the second mode $M_2$ to the previous case one notices that the polarization of the field has changed to the opposite --- $M_2$ is predominantly TM polarized. 

An increase in the order of the band does lead to enhancement of polarization mixing. However as we will see in more detail below, the substrate light-cone condition restricts the involvement of higher-order bands. We checked the polarization of the lowest six bands in several $k_{||}$ points and found that in all relevant cases certain symmetry, TE or TM, can be assigned to each band. Therefore, despite of highly asymmetric claddings (sapphire versus air) and relatively small refractive index contrast between PhCS ($n_{eff}\sim 2.1$) and the substrate ($n_{s}=1.78$), the mixing of the polarizations in the low-order photonic bands is limited. From now on we will ignore TM polarized bands. In the next section we will systematically study the effect of the structural parameters on the width of PBG in the spectrum of TE (see thick solid lines in Fig. \ref{cal}) modes. 

\section{IV Photonic band gap optimization}

\begin{figure}
\centerline{\rotatebox{-90}{\scalebox{0.29}{\includegraphics{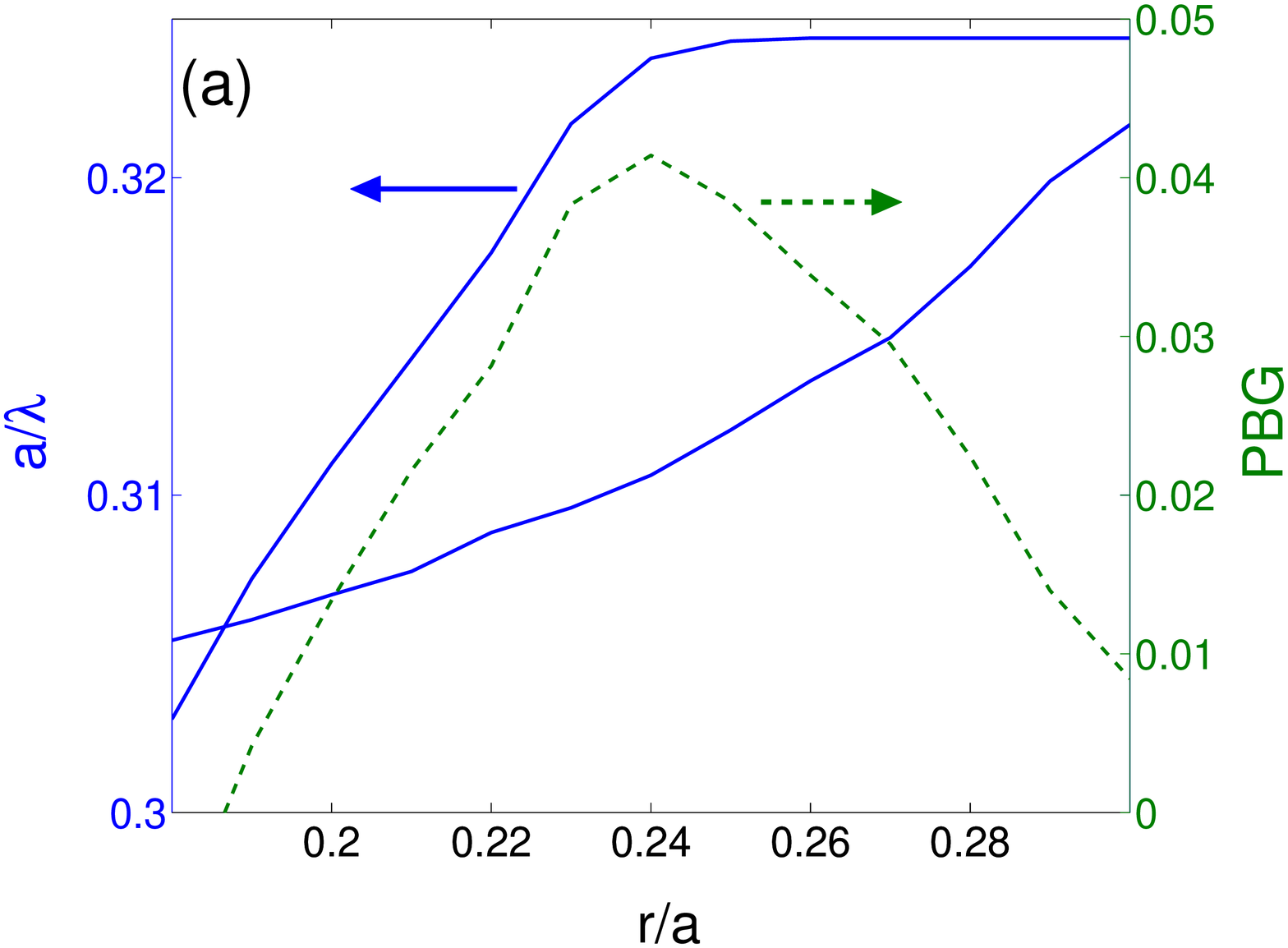}}}}
\centerline{\rotatebox{-90}{\scalebox{0.29}{\includegraphics{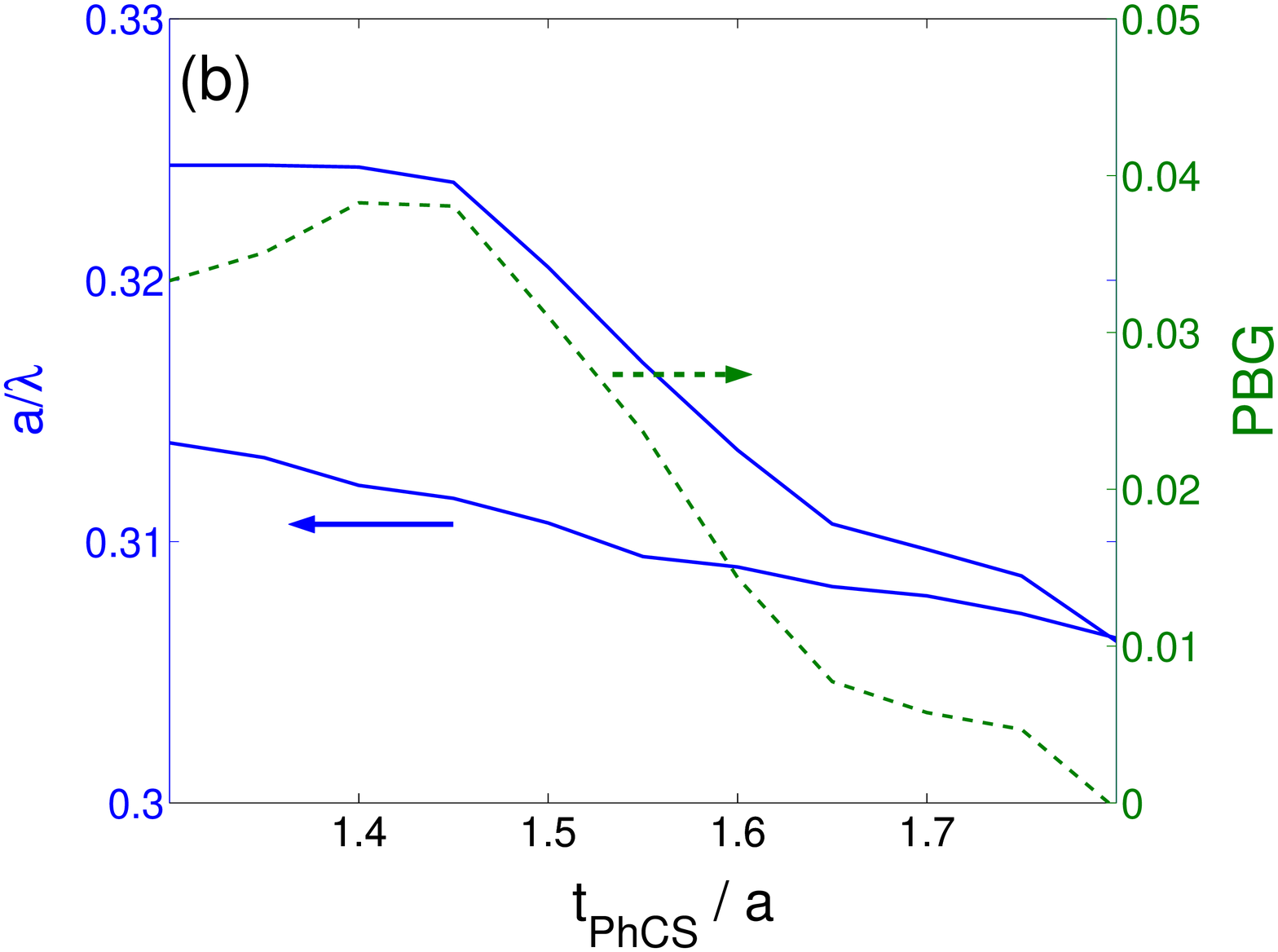}}}}
\centerline{\rotatebox{-90}{\scalebox{0.29}{\includegraphics{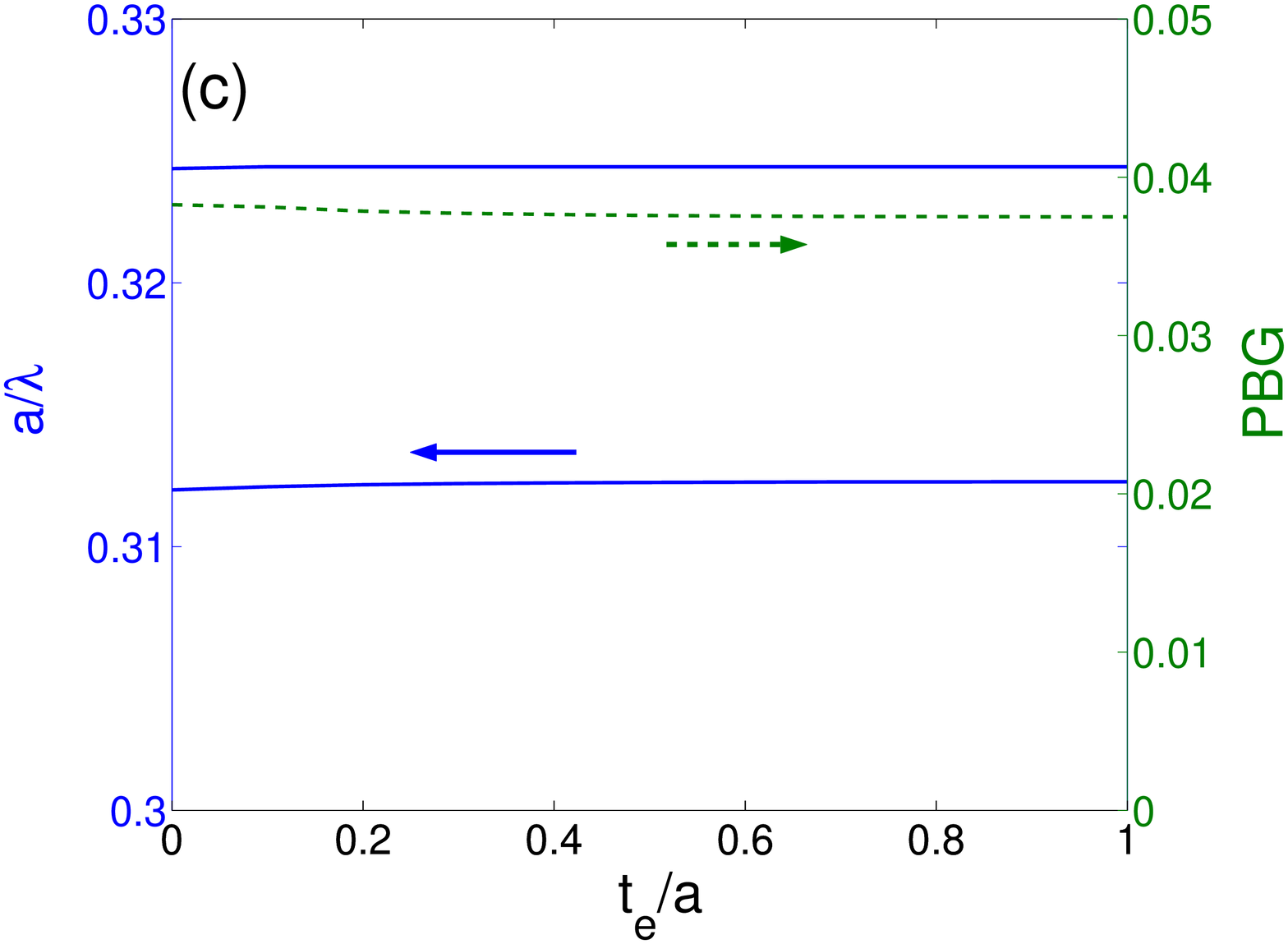}}}}
\caption{\label{gap_optimization} The position and relative size of photonic bandgap for PhCS with variable structural parameters. (a) $t_{PhCS}/a=1.4$, $t_e=0$; (b) $r/a=0.25$, $t_e=0$; (c) $r/a=0.25$, $t_{PhCS}/a=1.4$. $n_{ZnO}=2.35$ and $n_s=1.78$ were used in the calculation. The left vertical axes should be used with solid lines, which represent the edges of PBG. Dashed lines and the right vertical axes show the relative size of the gap.}
\end{figure}

We intend to optimize the following three parameters that can be controlled in the FIB etching\cite{our_apl}: (i) PhCS thickness, $t_{PhCS}$, (ii) air hole radius, $r$ (related to the filling fraction $f$) and etch depth, $t_e$. Due to linear nature of Maxwell equations, we normalize all parameters to the value of lattice constant, $a$. Our goal is to maximize the relative (normalized to the center frequency) PBG width. Once this is achieved, we can overlap PhCS gap with the emission spectrum of ZnO by choosing appropriate $a$. From the experimental prospective it is difficult to reproduce extremely small structural features, therefore we prefer to have PBG at high values of $a/\lambda$ if possible.

Fig. \ref{gap_optimization} shows the results of our simulation. In panel (a) of the figure, we plot the position and relative width of PhCS as we vary $r/a$ within the range $0.18-0.30$, and keep $t_{PhCS}/a=1.4$ and $t_e=0.0$. This corresponds to a change in the pattern filling fraction from $f=0.88$ to $f=0.67$. One can clearly see that there exists a PBG maximum at $r/a\simeq0.24$ (or $f\simeq0.79$). This effect has a clear physical interpretation: at large filling fractions (small $r/a$) the index contrast within PhCS is small, therefore band splitting, as well as PBG, is small. At small filling fractions, the wave guiding becomes poor, furthermore the decrease of the effective refractive index of the photonic layer leads to a shift of PBG as a whole to higher frequency. Eventually, at $r/a\simeq0.26$ the top of the gap is set not by the higher order guided mode but by $a/\lambda\simeq0.324$ at the light-cone when $k_{||}$ is at $M$-point (see Fig. \ref{cal}). In reality, above this frequency the guided modes will always be coupled to some leaky modes inside the light-cone. This phenomenon results in a plateau of the upper bandgap edge  for  $r/a=0.26-0.30$. Consequently, PBG disappears at $r/a>0.30$. For the air holes with such a large radius, the effective refractive index of PhCS $n_{eff}\simeq1.87$ approaches that of the substrate.

\begin{figure}
\centerline{\rotatebox{-90}{\scalebox{0.29}{\includegraphics{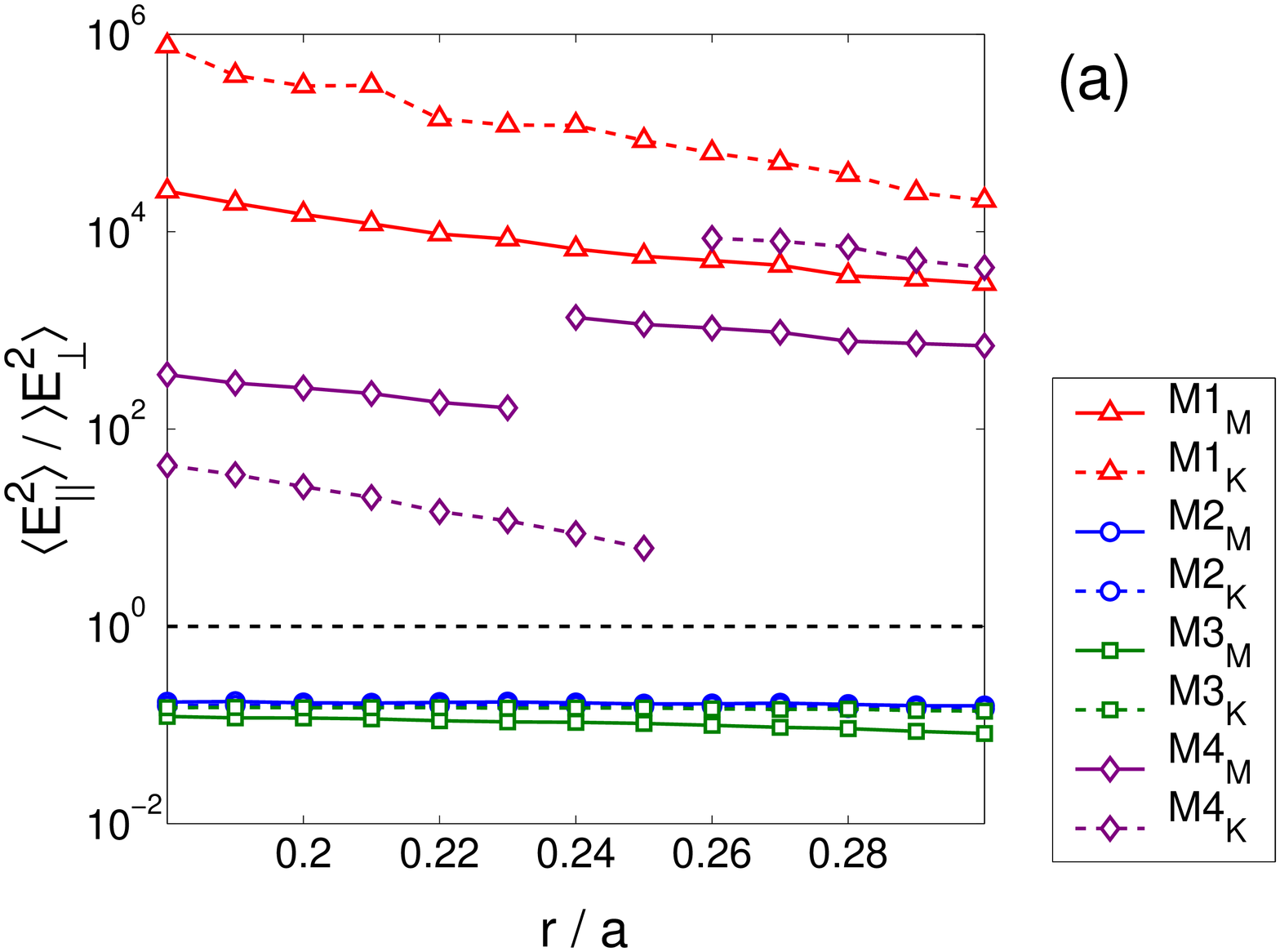}}}}
\centerline{\rotatebox{-90}{\scalebox{0.29}{\includegraphics{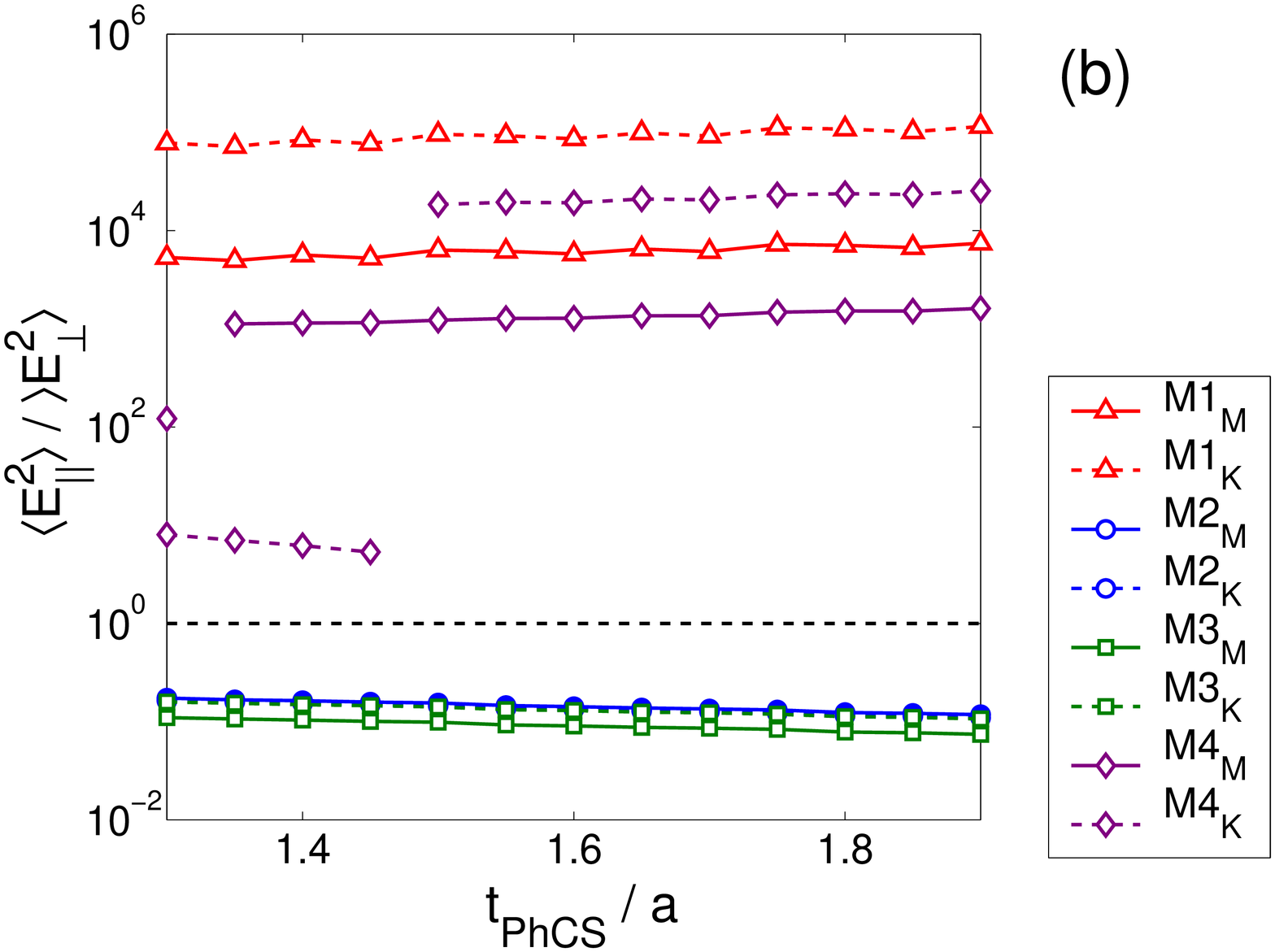}}}}
\centerline{\rotatebox{-90}{\scalebox{0.29}{\includegraphics{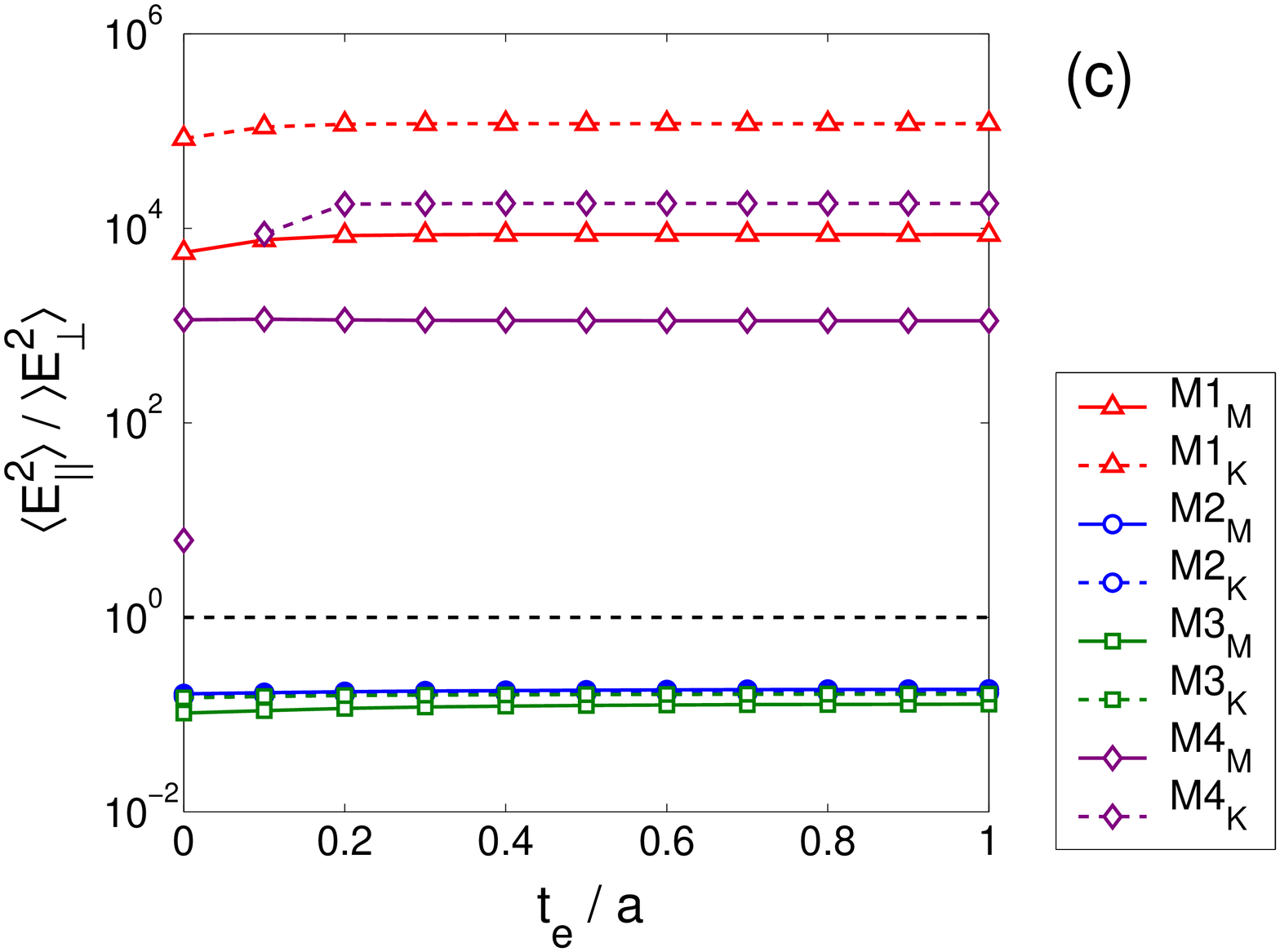}}}}
\caption{\label{polarization_optimization} The degree of polarization of the modes calculated for $k_{||}$ in $M$ (solid lines) and $K$ (dashed lines) k-points. The modes belonging to the lowest four bands are considered. Cases (a)-(c) correspond to those of Fig. \ref{gap_optimization}. $\langle ...\rangle$ denotes an integration over $x$ and $y$ coordinates. Discontinuities in the curves occur when the order of bands changes.}
\end{figure}

Fig. \ref{gap_optimization}(b) presents the effect of the photonic layer thickness, $t_{PhCS}$, on PBG. With the increase of $t_{PhCS}$ we also observe a maximum. For samples thinner than the correspondent wavelength, the eigenmodes of the system substantially ``spill'' into the air and (mostly) the substrate. Consequently, the effective index of PhCS becomes comparable to $n_s$ so that PBG lies at higher frequencies. The light-cone leads to the plateau of the upper bandgap edge as in the case of large $r/a$ in Fig. \ref{gap_optimization}(a). For very thick photonic layers, $t_{PhCS}/a>1.5$, the gap closes again. This occurs due to a stronger dependence of upper band edge frequency on the effective refractive index of PhCS. Indeed, as $n_{eff}$ is increased the band forming the upper edge of PBG develops the kinks close to the light-cone, that leads to the reduction of PBG as seen in Fig. \ref{gap_optimization}(b). As a result of these dependences, the size of the gap can be maximized at $t_{PhCS}/a\simeq1.45$.

The possibility of improving guiding properties of the photonic layer by extending the air holes into the substrate was discussed in Ref. \cite{benisty_etch_depth}. Indeed, removing a part of the substrate material lowers its refractive index and achieves a partial effect of ``undercutting''. However, the fact that PBG in our case lies close to the light-cone has a significant effect. An increase in $t_e$ results in the decrease of the effective refractive index experienced by a guided mode, because it extends into the substrate. Therefore the eigenfrequency of the mode increases and may enter the light-cone\cite{note}. This is exactly what happens in our system as shown in Fig. \ref{gap_optimization}c. As $t_e$ is increased the upper edge of PBG becomes defined by the position of the substrate light-cone at $M$ k-point (Fig. \ref{cal}), whereas the mode that defines the lower edge of PBG is more confined to PhCS and, therefore, is less sensitive to the effect of $t_e$. As a result, Fig. \ref{gap_optimization}(c) shows a weak dependence on the etch depth.

As we discussed in the previous section our calculation of the photonic band gap relies on the possibility of separating the photonic bands based on their polarization. To check the consistency of this assumption we calculated the ratios between in-plane and normal components of the electric fields for various structural parameters. In Fig. \ref{polarization_optimization}, we show the polarization ratios calculated for the lowest four bands at two $k_{||}$ vectors ($M$ and $K$ k-points). As the structural parameters are varied, the polarization of the modes is preserved. The most noticeable change occurs with the increase of $r/a$ in Fig. \ref{polarization_optimization}(a). This is due to significant modification of the band structure that occurs in this case. Notwithstanding, the modes still remain strongly polarized, justifying our methodology.

\section{Discussion and Conclusion}

In order to simulate the realistic structures to be realized in the experiment\cite{our_apl}, we modified super-cell technique\cite{Johnson:1999} within plane-wave expansion method for the photonic band structure calculation\cite{Johnson:2001}. We constructed a symmetric super-cell with the doubled size, Fig. \ref{geometry}. The induced degeneracy of eigenvalues provides a convenient way to verify the self-consistency of the calculation.

Substrate with high refractive index was expected to significantly mix the polarization of eigenmodes of PhCS. However, our calculations demonstrate that although the substrate induces asymmetry of wavefunctions (Fig. \ref{fields}), they remain strongly TM- or TE-polarized for low-order bands. High filling fraction air-hole-in-ZnO-matrix geometry can possess a complete PBG for TE bands, meanwhile, ZnO film with c-axis along the growth direction emits mainly into TE-polarized modes. This enabled us to build a UV PhCS laser\cite{our_apl}. In the photonic band structure calculation we find the optimum set of parameters for maximum PBG:
\begin{equation}
\frac{r}{a}\simeq 0.24\ \ \ \ \frac{t_{PhCS}}{a}\simeq 1.45\ \ \ \ \frac{t_e}{a}\simeq 0.0
\label{param}
\end{equation}
These parameters are significantly different from the typical parameters for IR PhCS. This is due to several factors: 
(i) presence of the sapphire substrate brakes vertical symmetry of PhCS; 
(ii) the refractive index contrast of ZnO/sapphire is lower than that of InP/air usually used in IR PhCS; 
(iii) to preserve guiding in the photonic layer the filling fraction and the thickness of ZnO PhCS needed to be significantly increased. This in turns explains the relatively small PBG of about $5\%$ that can be obtained in the ZnO PhCS on sapphire.

Overlapping the photonic bandgap with the emission spectrum of zinc oxide required precise control of the designed pattern with
\begin{equation}
a\simeq 123nm;t_{PhCS}\simeq 180nm;t_e\simeq 0nm;r\simeq 30nm.
\label{param}
\end{equation}
This has been achieved with the FIB etching technique. In order to avoid an additional damage to PhCS we did not attempt to remove the substrate, in other words, experimentally we preferred $t_e=0$. The maximum relative width of PBG that can be achieved via optimization is $5\%$. This is significantly smaller than what is being used in the membrane (air-bridge) geometry\cite{Painter:1999,park_electrically_pumped,bandedge_exp} or in the case of low refractive index substrate\cite{Chow:2000,phcs_on_low_index_substrate}. Narrow gap makes it difficult to align an intentionally introduced defect mode inside PBG. However, there are always some defect modes with frequencies inside PBG formed by small disorder unintentionally introduced in fabrication  process. These are the lasing modes observed in our experiment. 

This work was supported by the National Science Foundation under the grant no. ECS-0244457.

\end{document}